\documentclass[11pt,a4paper]{article}
\def\bea{\begin{eqnarray}}
\def\eea{\end{eqnarray}}

\def\beq{\begin{equation}}
\def\eeq{\end{equation}}
\def\ba{\beq\new\begin{array}{c}}
\def\ea{\end{array}\eeq}
\def\be{\ba}
\def\ee{\ea}
\newcommand{\Str}[2][\Big]{\ensuremath{\mathop{\rm STr}#1\{#2#1\}}}
\newcommand{\Pu}[2][\big]{\rm P#1[#2#1]}
\newcommand{\Cm}[2]{\ensuremath{\big[\Phi^{#1},\Phi^{#2}\big]}}
\newcommand{\re}[1]{(\ref{#1})}

\newcommand{\brane}{\mathop{\rm brane}}
\def\Dp{{\rm Dp}}
\def\l{\label}
\def\r{\ref}

\makeatletter
\newdimen\normalarrayskip % skip between lines
\newdimen\minarrayskip % minimal skip between lines
\normalarrayskip\baselineskip
\minarrayskip\jot
\newif\ifold \oldtrue \def\new{\oldfalse}
\def\arraymode{\ifold\relax\else\displaystyle\fi} % mode of array entries
\def\eqnumphantom{\phantom{(\theequation)}} % right phantom in eqnarray
\def\@arrayskip{\ifold\baselineskip\z@\lineskip\z@
\else
\baselineskip\minarrayskip\lineskip2\minarrayskip\fi}
\def\@arrayclassz{\ifcase \@lastchclass \@acolampacol \or
\@ampacol \or \or \or \@addamp \or
\@acolampacol \or \@firstampfalse \@acol \fi
\edef\@preamble{\@preamble
\ifcase \@chnum
\hfil$\relax\arraymode\@sharp$\hfil
\or $\relax\arraymode\@sharp$\hfil
\or \hfil$\relax\arraymode\@sharp$\fi}}
\def\@array[#1]#2{\setbox\@arstrutbox=\hbox{\vrule
height\arraystretch \ht\strutbox
depth\arraystretch \dp\strutbox
width\z@}\@mkpream{#2}\edef\@preamble{\halign
\noexpand\@halignto
\bgroup \tabskip\z@ \@arstrut \@preamble \tabskip\z@ \cr}%
\let\@startpbox\@@startpbox \let\@endpbox\@@endpbox
\if #1t\vtop \else \if#1b\vbox \else \vcenter \fi\fi
\bgroup \let\par\relax
\let\@sharp##\let\protect\relax
\@arrayskip\@preamble}
\def\eqnarray{\stepcounter{equation}%
\let\@currentlabel=\theequation
\global\@eqnswtrue
\global\@eqcnt\z@
\tabskip\@centering
\let\\=\@eqncr
$$%
\halign to \displaywidth\bgroup
\eqnumphantom\@eqnsel\hskip\@centering
$\displaystyle \tabskip\z@ {##}$%
\global\@eqcnt\@ne \hskip 2\arraycolsep
$\displaystyle\arraymode{##}$\hfil
\global\@eqcnt\tw@ \hskip 2\arraycolsep
$\displaystyle\tabskip\z@{##}$\hfil
\tabskip\@centering
&{##}\tabskip\z@\cr}
\begingroup\ifx\undefined\newsymbol \else\def\input#1 {\endgroup}\fi

\begin{document}
\setcounter{footnote}{1}
\def\thefootnote{\fnsymbol{footnote}}
\begin{center}
\hfill ITEP/TH-14/01\\
\hfill hep-th/0104034\\
\vspace{0.3in}
{\Large\bf Test for the  Myers-Chern-Simons Action
}
\end{center}
\centerline{{\large A. Alexandrov}\footnote
{ITEP, Moscow, Russia; e-mail: alex@gate.itep.ru}}
\setcounter{footnote}{0}
\def\thefootnote{\arabic{footnote}}
\bigskip

\abstract{\footnotesize We present a generalization of the infinitesimal gauge transformation
for nonabelian fields on the stack of branes up to the third order in $\Phi$. We test
the gauge invariance of the action
up to the fifth order in $\Phi$ for  $D$-instantons.  This substantiates the Myers
formula for the Chern-Simons term in action, which describes interaction with
the RR fields of $N$ coincident $\Dp$ branes.
}

\begin{center}
\rule{5cm}{1pt}
\end{center}

\bigskip
\section {Introduction}
$D$-branes are by definition hypersurfaces, where open strings can
end $\cite{Polchinski:1996na}$. Since closed strings can
interact with open strings, a separate $D$-brane can interact with closed strings.
$D$-branes are dynamical objects. Low-energy dynamics of the brane is described
by supersymmetric $U(1)$ gauge theory, bosonic fields of this theory
are adjoint scalars $\Phi$ and vector gauge field $A_a$. These fields can
be obtained from the ten dimensional open string theory by the dimensional
reduction. Scalars $\Phi$ describe transversal fluctuations of the brane.

 The brane interaction with closed strings can be described by the low
energy effective action. This action consists of two different parts. These are
the Dirac-Born-Infeld (DBI) term, describing Yang-Mills theory, and the Wess-Zumino term
(also known as the Chern-Simons action), governing the interaction of the
Ramond-Ramond (RR) fields with branes. The difference between them is that the DBI action does
not depend on the RR field and the Chern-Simons action is proportional to it. We
will consider only the latter one. But both of them, apart from being
invariant under the $U(1)$ gauge transformation
\be
A\rightarrow A+\Pu{d\epsilon},
\l{1}
\ee
with one scalar arbitrary gauge parameter $\epsilon$, are also invariant under some other
transformations, for example, a more
general transformation

\be
A\rightarrow A+\Pu{\Lambda},\\
B\rightarrow B-d\Lambda,
\l{2}
\ee
with a 1-form gauge parameter $\Lambda$. Obviously, transformation \re{1}
is a particular case
of \re{2} with exact $\Lambda$, i.e. $\Lambda=d\epsilon$ for some
$\epsilon$.

If there are N coincident $\Dp$-branes, $U(1)$ groups extend to the
$U(N)$\cite{Witten:1996im}.
The action in this case is a nontrivial generalization of the Abelian action.
R.C.Myers in his work \cite{Myers:1999ps} derived this action for a  particular
case and proposed a general
formula. The goal of this paper is to test the Myers action. As we
know from string theory, the correct action is inevitably gauge invariant.
Therefore
we check the gauge invariance of the Myers action under gauge transformations like
\re{2}. The problem is that the generalization of \re{2} for the case
of multiple $D$-branes is still unknown. We derive it up to the third order in the scalar
field $\Phi$. The proper transformation looks as follows
\be
\Phi^k\rightarrow\Phi^k+\big[\Phi^k,\Lambda_i(0)\Phi^i+
\frac{1}{2}\partial_j\Lambda_i(0)\Phi^i\Phi^j\big]+\\
+\partial_j\Lambda_i(0)\big(\frac{1}{2}\big[\Phi^j,
\Phi^k\big]\Phi^i+\frac{1}{2}\big[\Phi^k,\Phi^i\big]\Phi^j+
\beta\big[\big[\Phi^i,\Phi^j\big],\Phi^k\big]\big)+O(\Phi^4),
\l{5}
\ee
and the coefficient $\beta$ can not be determined from our consideration.
Possibly, it can be determined from the consideration of higher than fifth
orders in $\Phi$ in the expansion of the action.

The paper is organized as follows: in section 2 we describe our notations
and review properties of the Abelian Chern-Simons action. In section 3 we
generalize all notations for the stack of branes and present the gauge
transformation for the exact gauge parameter. In section 4 we derive the gauge
transformation for general gauge parameter up to the third order in $\Phi$, analysing
the action for the
simplest case of D-instantons, which we expand up to the fifth order in $\Phi$.
We test the transformation of the same form in the more
difficult case of $D2$-branes in section 5. In Appendix we show that
the simplest generalization of the Abelian action is not gauge invariant,
and therefore, is not correct.

\section {Abelian Chern-Simons Action}

 In the type IIA and IIB theories massless modes of the open string sector are
the gravitational field $G$, Calb-Ramond 2-form field $B$, Ramond-Ramond
poli-form $C$ and dilaton scalar $\varphi$. Respectively, in the
type IIA case the RR field is the sum of odd forms and in the type IIB -- the sum
of even forms.
The massless mode
of open strings is a one-form gauge field. If we consider a brane,
which breaks translational invariance in the bulk,
some components of this one--form become scalar fields $\Phi$ on the brane,
and other induce the gauge field on the brane with field strength $F$.

The low energy action for the brane is described by the sum of the
Dirac-Born-Infeld and Chern-Simons actions
\footnote
{Our notation slightly differs from that of R.C.Myers, namely, we set $T_p\rightarrow1$,
$\mu_p\rightarrow1$, $iA\rightarrow A$ and $i\lambda\rightarrow 1$. We also have
the different sign in front of $F$.}
\be
S=S_{DBI}+S_{CS}=\\ =-\int_{W_{p+1}}d^{p+1}\sigma\Big(e^{-\varphi}\sqrt{
-\det\bigl(\Pu{G+B}_{ab}+F_{ab}\bigr)}\Big)+ \int_{W_{p+1}}
\Pu{Ce^{B+F} },
\l{ac}
\ee
where $W_{p+1}$ is the p+1 dimensional brane world-volume, and $\Pu{\ldots}$ is
a pull-back operation, which we determine later. We will split the bulk
co-ordinates ($X^\mu, \mu=0,\ldots n$, $n$ is the number of bulk dimensions)
into two parts: co-ordinates along
the brane, which coincide with the co-ordinate system on the brane  ($x^a=X^a,
$ $a=0,\ldots p$) and those orthogonal to it ($X^i$, $i=p+1,\ldots,n$). This means we use the so called
"static gauge". The position
of the brane in the bulk is described by the set of the scalar fields
$\Phi^i(x^a)$.
 From the brane
point of view, closed string fields are the fields on the brane,
depending on scalars $\Phi$. We consider them as series in $\Phi$, for
example

\be
B_{\mu\nu}(X^a,X^i)\Big|_{X=X_{\brane}}=B_{\mu\nu}(x^a,0)+
\partial_iB_{\mu\nu}(x^a,0)\Phi^i(x^a)+\ldots
\ee
From now on we omit the explicit dependence of all the fields on
co-ordinates on the brane $x^a$.

Since $F$ is a form on the brane, and the closed string fields $B$ and $C$ are forms in
the bulk, there are two different types of forms: forms in the bulk and forms on
the world-volume. Respectively, we have distinctive differentials $d^{\brane}$ and
$d$, and different wedge products $\wedge^{\brane}$ and $\wedge$.
We are bound to convert bulk tensors into world-volume tensors. To this end we
apply the pull-back operation $P\big[\ldots\big]$. The pull-back acts, for
example, on the 3-form as follows
\be
\Pu{C^{(3)}}_{abc}=
C^{(3)}_{\mu\nu\chi}(X)\partial_aX^\mu\partial_bX^\nu\partial_cX^\chi\Big|_{X=X_{\brane}}=\\
=C^{(3)}_{abc}(\Phi)+C^{(3)}_{abi}(\Phi)\partial_c\Phi^i+C^{(3)}_{aic}(\Phi)\partial_b\Phi^i
+C^{(3)}_{ibc}(\Phi)\partial_a\Phi^i+
C^{(3)}_{aij}(\Phi)\partial_b\Phi^i\partial_c\Phi^j+\\
+C^{(3)}_{ibj}(\Phi)\partial_a\Phi^i\partial_c\Phi^j+
C^{(3)}_{ijc}(\Phi)\partial_a\Phi^i\partial_b\Phi^j+
C^{(3)}_{ijk}(\Phi)\partial_a\Phi^i\partial_b\Phi^j\partial_c\Phi^k.
\ee
The pull-back is the homomorphism, that is

\be
\Pu{U\wedge W}=\Pu{U}\wedge^{\brane} \Pu{W}.
\ee
This means we can rewrite the Chern-Simons term in another manner

\be
S=\int_{W_{p+1}} \Pu {C}e^{\Pu{B}+F}.
\l{ac1}
\ee
The Chern-Simons action is invariant with respect to the gauge transformations of two
different types $\cite{Alekseev:2000ch}$:

1) Ramond-Ramond field transformations
\be
C\rightarrow C+ e^{-B}d\epsilon,
\ee

2) transformation, leaving combination $P\big[B\big]+F$ invariant
\be
A\rightarrow A+\Pu{\Lambda},\\
B\rightarrow B-d\Lambda.
\l{tr}
\ee
This latter transformation does not change the combination $P\big[B\big]+F$, since
$F=d^{\brane}A$
and $$d^{\brane}\Pu{\ldots}=\Pu{d\ldots}.$$ We investigate its
generalization for the nonabelian case.
If $\Lambda=d\epsilon$, transformation \re{tr} is just $U(1)$
transformation \re{1}.

\section {Nonabelian Chern-Simons Action}
If we consider a stack of branes, the situation is more difficult. Now the gauge theory
on the stack of branes is nonabelian, and the action can acquire new features. For example,
the theory celebrates the so called "dielectric effect", i.e. the interaction of branes
with the RR field of higher than $p+1$ dimension emerges \cite{Myers:1999ps}. Now we
want to generalize the action \re{ac1}. In order to determine the action for the
nonabelian case, we must introduce trace with some ordering and determine $\Pu{\ldots}$.
 We also can add
some terms vanishing in the Abelian case. The generalization of the
pull-back is straightforward -- we must substitute the covariant derivatives for
the usual ones

\be
\partial_a\Phi^i\rightarrow D_a\Phi^i=\partial_a\Phi^i+\big[A_a,\Phi^i\big]
\ee
The most natural ordering is $\Str{\ldots}$ (see \cite{Tseytlin:1997cs}).
This is a symmetric trace, i.e. it is symmetrical in $\Phi^i$, $D_a\Phi^i$ and
$F_{ab}$.
With this prescription pull-back operation remains homomorphism
\be
\Str{\Pu{A}\wedge^{brane}\Pu{B}}=\Str{\Pu{A\wedge B}}.
\l{111}
\ee
Thus, the most natural generalization of the action \re{ac1} is
\be
S=\int_{W_{p+1}} \Str{\Pu {C}e^{\Pu[]{B}+F}}.
\l{mng}
\ee
However, it is impossible to recover the gauge invariance for this action (see
the Appendix). Hence, this action is not correct. Using T-duality, R.C.Myers
proposed another action for the stack of branes \cite{Myers:1999ps}
\be
S=\int_{W_{p+1}}\Str{\Pu{e^{i_\Phi i_\Phi}Ce^{B+F}}}
\l{acs}
\ee

Our aim here is
to find the generalization of transformation \re{tr}.
The generalization of the $U(1)$ transformations \re{1} is
the infinitesimal $U(N)$ transformations
\be
A\rightarrow A+\Pu{d\epsilon}+\big[A,\epsilon\big],\\
\Phi\rightarrow \Phi+\big[\Phi,\epsilon\big],
\l{imvn}
\ee
since the finite $U(N)$ transformations are
\be
A\rightarrow G\partial G^{-1}+GAG^{-1},\\
\Phi\rightarrow G\Phi G^{-1}.
\ee
One obtains \re{imvn} for $G=e^{-\epsilon}$. Thus, for the exact
$\Lambda=d\epsilon$,
one knows the gauge transformation
\be
A_a\rightarrow A_a+\Lambda_a(\Phi)+\Lambda_i(\Phi)\partial_a\Phi^i+\\
+\big[A_a,\Lambda_i(0)\Phi^i+
\frac{1}{2}\partial_j\Lambda_i(0)\Phi^i\Phi^j+O(\Phi^3)\big],\\
\Phi^k\rightarrow \Phi^k+\big[\Phi^k,\Lambda_i(0)\Phi^i+
\frac{1}{2}\partial_j\Lambda_i(0)\Phi^i\Phi^j\big]+O(\Phi^4).
\l{nt}
\ee
Now we determine gauge transformations for the generic $\Lambda$. We
perform our calculations in the case of $D$-instantons.
\section {$D$-instantons}
The  action \re{acs} has the simplest form in the case of D(-1) branes. This action
is
\be
S=\Str{e^{\rm i_\Phi i_\Phi}Ce^{B}}.
\ee
Up to fifth power in $\Phi$, the action is of the form

\be
S=\Str{ C^{(0)}+{\rm i_\Phi i_\Phi} \big(C^{(0)}B+C^{(2)}\big)+\frac{1}{2}\big({\rm i_\Phi i_\Phi}\big)^2
\big(\frac{1}{2}C^{(0)}B^2+C^{(2)}B+C^{(4)}\big)}.
\ee
We consider infinitesimal gauge transformation as a series in $\Phi$ and, in general, it
derivatives
\be
\Phi \rightarrow \Phi +\delta_{(2)}\Phi+\delta_{(3)}\Phi+\ldots
\ee
The gauge invariance imposes equations for $\delta_{(3)}\Phi$, simplest of
them are the follows
\beq\left\{\begin{array}{l}
 \Str{\partial_kC^{(0)}(\Phi)\delta_{(3)}\Phi^k-C^{(0)}(\Phi)\Cm{i}{j}\partial_j\Lambda_i0)}=0,\\
 \Str{\frac{1}{2}\Cm{i}{j}\partial_k\big(C^{(0)}(\Phi)B_{ji}(\Phi)\big)\delta_{(3)}\Phi^k
 +\big[\Phi^i,\delta_{(3)}\Phi^j\big]C^{(0)}(\Phi)B_{ji}(\Phi)-\\
 -\frac{1}{8}\Cm{i}{j}\Cm{m}{n}C^{(0)}(\Phi)\big(B(\Phi)\wedge
 d\Lambda(0)\big)_{nmji}}=0,\\
 \Str{\frac{1}{2}\Cm{i}{j}\partial_kC^{(2)}_{ji}(\Phi)\delta_{(3)}\Phi^k
 +\big[\Phi^i,\delta_{(3)}\Phi^j\big]C_{ji}^{(2)}(\Phi)+\\
 -\frac{1}{8}\Cm{i}{j}\Cm{m}{n}\big(C^{(2)}(\Phi)\wedge
 d\Lambda(0)\big)_{nmji}}=0.
\end{array}\right.\eeq
This gives a system determining $\delta_{(3)}\Phi$
\beq\left\{\begin{array}{l}
\partial_kC(0)\Str{\delta_{(3)}\Phi^k}-\partial_kC(0)
\partial_j\Lambda_i(0)\Str{\big[\Phi^i,\Phi^j\big]\Phi^k}=0\\
\partial_k\partial_mC(0)\Str{\delta_{(3)}\Phi^k\Phi^m}-
\frac{\partial_k\partial_mC(0)}{2!}\partial_j\Lambda_i(0)
\Str{\big[\Phi^i,\Phi^j\big]\Phi^k\Phi^m}=0\\
\frac{\partial_k\partial_m\partial_nC(0)}{2!}\Str{\delta_{(3)}\Phi^k\Phi^m\Phi^n}-
\frac{\partial_k\partial_m\partial_nC(0)}{3!}\partial_j\Lambda_i(0)
\Str{\big[\Phi^i,\Phi^j\big]\Phi^k\Phi^m\Phi^n}=0\\
\Str{\frac{1}{2}\Cm{i}{j}\partial_k\big(C^{(0)}(0)B_{ji}(0)\big)\delta_{(3)}\Phi^k
+\big[\Phi^i,\delta_{(3)}\Phi^j\big]\partial_k\big(C^{(0)}(0)B_{ji}(0)\big)\Phi^k-\\
-\frac{1}{8}\Cm{i}{j}\Cm{m}{n}\big(\partial_k\big(C^{(0)}(0)B(0)\big)\wedge
d\Lambda(0)\big)_{nmji}\Phi^k}=0,\\
\Str{\frac{1}{2}\Cm{i}{j}\partial_kC^{(2)}_{ji}(0)\delta_{(3)}\Phi^k
+\big[\Phi^i,\delta_{(3)}\Phi^j\big]\partial_kC_{ji}^{(2)}(0)\Phi^k-\\
-\frac{1}{8}\Cm{i}{j}\Cm{m}{n}\big(\partial_kC^{(2)}(0)\wedge
d\Lambda(0)\big)_{nmji}\Phi^k}=0
\l{smg}
\end{array}\right.\eeq
The most general form of the $\delta_{(3)}\Phi$ is
\be
\delta_{(3)}\Phi^k=\big(\alpha_1\big[\Phi^j,\Phi^k\big]\Phi^i+\alpha_2\big[\Phi^i,\Phi^j\big]\Phi^k+
\alpha_3\big[\Phi^k,\Phi^i\big]\Phi^j+\\
+\beta_1\big[\big[\Phi^j,\Phi^k\big],\Phi^i\big]+
\beta_2\big[\big[\Phi^i,\Phi^j\big],\Phi^k\big]+\beta_3\big[\big[\Phi^k,\Phi^i\big],\Phi^j\big]+\\
+\gamma_1\big[\big\{\Phi^j,\Phi^k\big\},\Phi^i\big]+
\gamma_2\big[\big\{\Phi^i,\Phi^j\big\},\Phi^k\big]+
\gamma_3\big[\big\{\Phi^k,\Phi^i\big\},\Phi^j\big]\big)\partial_j\Lambda_i(0),
\l{mg}
\ee
where $\big\{\Phi^k,\Phi^j\big\}$ is anticommutator
\be
\big\{\Phi^k,\Phi^j\big\}\equiv\Phi^k\Phi^j+\Phi^j\Phi^k.
\ee
We can put some $\beta$'s to zero, using the Jacobi identity
\be
\big[\big[\Phi^j,\Phi^k\big],\Phi^i\big]+
\big[\big[\Phi^i,\Phi^j\big],\Phi^k\big]+\big[\big[\Phi^k,\Phi^i\big],\Phi^j\big]\equiv0,
\ee
and put some $\gamma$'s to zero, using the identity
\be
\big[\big\{\Phi^j,\Phi^k\big\},\Phi^i\big]+
\big[\big\{\Phi^i,\Phi^j\big\},\Phi^k\big]+
\big[\big\{\Phi^k,\Phi^i\big\},\Phi^j\big]\equiv0.
\ee
Substituting \re{mg} into \re{smg}, one obtains the restrictions on
coefficients $\alpha$, $\beta$ and $\gamma$
\beq\left\{\begin{array}{l}
\alpha_1+\alpha_3=1\\
\beta_1=\beta_3=\alpha_2=\gamma_1=\gamma_3=0\\
\end{array}\right.\eeq

This system with the requirement that the transformation is the generalization
of the transformation \re{nt} for $d\Lambda\neq0$ gives
\beq\left\{\begin{array}{l}
\gamma_1=\gamma_3=\alpha_2=\beta_1=\beta_3=0\\
\alpha_1=\alpha_3=\frac{1}{2}\\
\gamma_2=-\frac{1}{4}
\end{array}\right.\eeq
So we determined all but one coefficients. Possibly, $\beta_2$ can be determined from
the higher power expansion of the action. Now we test this transformation
with the $D2$-branes action.
\section {$D$2-branes}
We test the gauge invariance of the action for the $D2$-branes up to third
power in $\Phi$. Consider the field configuration with $B=0$, $A=0$, $C^{(1)}=0$ and $C^{(5)}=0$. The action is
\be
S=\int_{W_3}\Str{\Pu[\Big]{C^{(3)}+i_{\Phi}i_{\Phi}\big(C^{(3)}\wedge\big(B+F\big)\big)}}.
\ee
The variation of the
action due to the "dielectric" effect is as follows

\be
\delta_S S=\delta \int_{W_3}  \Str{P\Big[{\rm i}_\Phi {\rm i}_\Phi \Big(C^{(3)}\wedge\big(B+F\big)\Big)\Big]}=\\
=\int_{W_3}\Str{{\rm i}_{\Phi^j}{\rm i}_{\Phi^i}\Big(C\wedge\delta\big(B+F\big)\Big)_{ij\mu\nu\chi}
D_aX^\mu D_bX^\nu D_cX^\chi}\Big|_{X=X_{\brane}}\epsilon^{abc}d^3V=\\
=\int_{W_3}  \Str{\frac{1}{2}\Cm{i}{j}\Big(3C_{ija}
\bigl(d\Lambda_{bc}-dP[\Lambda]_{bc}\bigr)+C_{abc}d\Lambda_{ij}+
6C_{jbc}d\Lambda_{ai}\Big)+\\
+3\Big(2\big[\Phi^i,\Phi^j\big]C_{jkc}d\Lambda_{ib}\partial_a\Phi^k+
\frac{1}{2}\big[\Phi^i,\Phi^k\big]C_{jbc}d\Lambda_{ik}\partial_a\Phi^j+\\
+\big[\Phi^j,\Phi^i\big]C_{jbc}d\Lambda_{ik}\partial_a\Phi^k+
\big[\Phi^j,\Phi^k\big]C_{jkc}d\Lambda_{bi}\partial_a\Phi^i
\Big)}\epsilon^{abc}d^3V.
\ee
The gauge invariance of the actions puts the condition (see the Appendix)
\be
\delta S =\delta_A S+\delta_\Phi S+\delta_S S =0,
\ee
which gives the system
\beq\left\{\begin{array}{l}
\Str{\partial_kC_{abc}(0)\delta_{(3)}\Phi^k+
\frac{1}{2}\big[\Phi^i,\Phi^j\big]\partial_kC_{abc}(0)
\big(\partial_i\Lambda_j(0)-\partial_j\Lambda_i(0)\big)\Phi^k}\epsilon^{abc}=0,\\
\Str{C_{jbc}(0)\partial_a\delta_{(3)}\Phi^j+
\big[\Phi^i,\Phi^j\big]\Big(C_{jbc}(0)\Phi^k\partial_k\partial_a\Lambda_i(0)\Big)+\\
~~~~~~+\big[\Phi^i,\Phi^k\big]\Big(C_{jbc}(0)\partial_a\Phi^j\partial_i\Lambda_k(0)\Big)+\\
~~~~~~+\big[\Phi^j,\Phi^i\big]\Big(C_{jbc}(0)\partial_a\Phi^k\big(\partial_i\Lambda_k(0)-\partial_k\Lambda_i(0)\big)\Big)
}\epsilon^{abc}=0,\\
\Str{\partial_k\partial_mC_{abc}(0)\Phi^k\delta_{(2)}\Phi^m}\epsilon^{abc}=0,\\
\Str{\partial_kC_{jbc}(0)
\Bigl(
\delta_{(2)}\Phi^k\partial_a\Phi^j+\Phi^k\partial_a\delta_{(2)}\Phi^j\Bigr)+\\
~~~~~~+\partial_kC_{jbc}(0)\Big[\Lambda_i(0)\partial_a
\Phi^i+\partial_i\Lambda_a(0)\Phi^i,\Phi^j\Big]\Phi^k+\\
~~~~~~+\big[\Phi^i,\Phi^j\big]\Phi^k\partial_kC_{jbc}(0)
\big(\partial_a\Lambda_i(0)-\partial_i\Lambda_a(0)\big)}\epsilon^{abc}=0,\\
\Str{C_{jkc}(0)\partial_a\delta_{(2)}\Phi^j\partial_b\Phi^k+
C_{jkc}(0)\partial_b \Phi^k\Big[\Lambda_i(0)\partial_a
\Phi^i+\partial_i\Lambda_a(0)\Phi^i,\Phi^j\Big]\\
~~~~~~+\frac{1}{2}\big[\Phi^i,\Phi^j\big]\Bigl(
C_{ijc}(\partial_a\Lambda_i(0)\partial_b\Phi^i+\partial_i\Lambda_b(0)\partial_a\Phi^i\big)+\\
~~~~~~\big[\Phi^i,\Phi^j\big]C_{jkc}\big(\partial_i\Lambda_b(0)-\partial_b\Lambda_i(0)\big)\partial_a\Phi^k+\\
~~~~~~\frac{1}{2}\big[\Phi^k,\Phi^j\big]C_{jkc}(0)\big(\partial_i\Lambda_b(0)-\partial_b\Lambda_i(0)\big)
\partial_a\Phi^k\Bigr)}\epsilon^{abc}=0.
\end{array}\right.\eeq
It is easy to check that the gauge transformations found in section 4 solve these
equations.
\section {Conclusions}
We obtained the manifest expression for the gauge transformation for arbitrary gauge
parameter up to the third power in $\Phi$. The fact that for the
Myers-Chern-Simons action it is possible to restore gauge invariance,
supports its form. However, our consideration does not allow one to restore it unambiguously.
It would be interesting to test the same gauge invariance with the nonabelian
DBI action.
\section {Acknowledgements}
Author is grateful to  E. Akhmedov and A. Mironov for careful reading the manuscript and useful discussions,
to A. Dymarsky, S. Klevtsov and D. Melnikov for
discussions and especially to A. Morozov for initiating this work and
support. This work was partly supported by the Russian President's grant
00-15-99296, $RFBR$ grant 00-02-16477 and grant $INTAS$--00--334.
\section {Appendix}
It is easy to see that in contrast to the Myers action \re{acs}, it is impossible to construct the gauge
transformations, generalizing \re{2}, for the action \re{mng} . We shall examine the case of 2-branes in
the case $B=0$, $C^{(1)}=0$ and $A=0$ up to the third power in $\Phi$.
In this case

\be
S=\int_{W_3}\Str{\Pu{C^{(3)}(\Phi)}},
\l{32}
\ee
and
\be
\delta S =\delta_A S+\delta_\Phi S
\ee
does not vanish. Implicitly we have
\be
\delta_A S=
3\int_{W_3} \Str{
[\Lambda_i\partial_a \Phi^i+\partial_i\Lambda_a\Phi^i,\Phi^j]
\left(2C^{(3)}_{j k c}(0)\partial_b \Phi^k+
\partial_kC^{(3)}_{j b c}(0)\Phi^k\right)}
\epsilon^{a b c}d^3V,
\l{dels}
\ee
and
\be
\delta_\Phi S=
\int_{W_3} \mathop{\rm STr}
\biggl\{\partial_kC_{abc}(0)\delta_{(3)}\Phi^k+
\partial_k\partial_mC_{abc}(0)\Phi^k\delta_{(2)}\Phi^m+3C_{jbc}(0)\partial_a\delta_{(3)}\Phi^j+\\
+3\partial_kC_{jbc}(0)
\Bigl( \delta_{(2)}\Phi^k\partial_a\Phi^j+\Phi^k\partial_a\delta_{(2)}\Phi^j\Bigr)+
6C_{jkc}(0)\partial_a\delta_{(2)}\Phi^j\partial_b\Phi^k
\biggr\}\epsilon^{abc}d^3V.
\ee
If the action \re{32} was correct, we could chose $\delta_{(2)}\Phi$ and
$\delta_{(3)}\Phi$ to cancel the full $\delta S$

\be
\delta S =\delta_A S+\delta_\Phi S =0
\ee
However, we want to show, there is no proper $\delta_{(2)}\Phi$, and the action \re{32}
is not exact.  As $C(0)$
and $\partial C(0)$ are independent coefficients, we have the sets of equations
for determination of $\delta_{(3)}\Phi$ and $\delta_{(2)}\Phi$

\beq\left\{\begin{array}{l}
\Str{\partial_kC_{abc}(0)\delta_{(3)}\Phi^k}\epsilon^{abc}=0,\\
\Str{C_{jbc}(0)\partial_a\delta_{(3)}\Phi^j}\epsilon^{abc}=0.
\end{array}\right.\eeq
\beq\left\{\begin{array}{l}
\Str{\partial_k\partial_mC_{abc}(0)\Phi^k\delta_{(2)}\Phi^m}\epsilon^{abc}=0,    \\
\Str{\partial_kC_{jbc}(0)\Bigl(
\delta_{(2)}\Phi^k\partial_a\Phi^j+\Phi^k\partial_a\delta_{(2)}\Phi^j\Bigr)+~~~~~~~~~(*)\\
\partial_kC_{jbc}(0)\Big[\Lambda_i(0)\partial_a \Phi^i+\partial_i\Lambda_a(0)\Phi^i,\Phi^j\Big]\Phi^k
}\epsilon^{abc}=0,\\
\Str{C_{jkc}(0)\partial_a\delta_{(2)}\Phi^j\partial_b\Phi^k+\\
C_{jkc}(0)\partial_b \Phi^k\Big[\Lambda_i(0)\partial_a
\Phi^i+\partial_i\Lambda_a(0)\Phi^i,\Phi^j\Big]
}\epsilon^{abc}=0.~(**)
\l{sys}
\end{array}\right.\eeq
Now we show that it is impossible to
find any appropriate $\delta_{(2)}\Phi$.

The most general form of $\delta_{(2)}\Phi$ is

\be
\delta_{(2)}\Phi^j=\sum_{N\leq M}L_{pq}^{j;b_1\ldots b_N;c_1\ldots c_M}
\Bigl[\partial_{b_1}\ldots\partial_{b_N}\Phi^p,\partial_{c_1}\ldots\partial_{c_M}\Phi^q\Bigr],
\ee
with $L$ being commutative coefficients.
From the system ($\r{sys}$) one gets
\be
\delta_{kj}\Str{\Phi^k\delta_{(2)}\Phi^j}=0
\l{22}
\ee
This gives the condition

\be
L_{pq}^{j;b_1\ldots b_N;c_1\ldots c_M}=0,
\ee
if $N>0$. So one can rewrite the equation \re{22} in the following form
\be
\delta_{kj}\Str{\Phi^k\Big(L^j_{pq}\Big[\Phi^p,\Phi^q\Big]+
\sum_{M>0}L_{pq}^{j;c_1\ldots c_M}
\Bigl[\Phi^p,\partial_{c_1}\ldots\partial_{c_M}\Phi^q\Bigr]\Big)}=0.
\ee
Thus, we have a system for identification of $\delta_{(2)}\Phi$ and
\be
L_{pq}^{j;c_1\ldots c_M}=\delta^j_pL_q^{c_1\ldots c_M}
\ee
for $M\geq0$. We can substitute
\be
\delta_{(2)}\Phi^j=\Big[\Phi^j,\sum_{M=0}^\infty L_q^{c_1\ldots
c_M}\partial_{c_1}\ldots\partial_{c_M}\Phi^q\Big]=\Big[\Phi^j,L\Phi\Big],
\ee

\be
\partial_a\delta_{(2)}\Phi^j=
\Big[\partial_a\Phi^j,L\Phi\Big]+
\Big[\Phi^j,\partial_a\big(L\Phi\big)\Big],
\ee
into system \re{sys} and find $L$ from the equation (*) of \re{sys}
\be
\Str{\Bigl(
\delta_{(2)}\Phi^k\partial_a\Phi^j+\Phi^k\partial_a\delta_{(2)}\Phi^j\Bigr)+
\Phi^k\Big[\Lambda_i(0)\partial_a \Phi^i+\partial_i\Lambda_a(0)\Phi^i,\Phi^j\Big]}=\\
\Str{
\Big[\Phi^k,L\Phi\Big]\partial_a\Phi^j+\Phi^k\Bigl(\Big[\partial_a\Phi^j,L\Phi\Big]+
\Big[\Phi^j,\partial_aL\Phi\Big]\Bigr)+\\
\Phi^k\Big[\Lambda_i(0)\partial_a \Phi^i+\partial_i\Lambda_a(0)\Phi^i,\Phi^j\Big]}=0.
\ee
Since $\Phi$'s are arbitrary, the trace vanishes only if
\be
\Big[L\Phi,\partial_a\Phi^j\Big]+\Big[\partial_a\Phi^j,L\Phi\Big]+
\Big[\Phi^j,\partial_aL\Phi\Big]+
\Big[\Lambda_i(0)\partial_a
\Phi^i+\partial_i\Lambda_a(0)\Phi^i,\Phi^j\Big]=0.
\ee
In a similar manner
\be
\partial_aL\Phi-\Lambda_i(0)\partial_a\Phi^i-\partial_i\Lambda_a(0)\Phi^i=0.
\ee
Equation (**) of the system \re{sys} gives the same condition. As $\Phi$'s and their
derivatives are independent matrices, one has conditions

\beq\left\{\begin{array}{l}
\partial_aL_i-\partial_i\Lambda_a(0)=0,\\
\big(\partial_aL_i^c\big)\partial_c\Phi^i+L_i\partial_a\Phi^i-\Lambda_i(0)\partial_a\Phi^i=0.
\l{f}
\end{array}\right.\eeq
One observes, that
\be
\partial_aL_i^c=\delta_a^cN_i
\ee
for some $N_i$.
The only possibility is $N_i=const$, but since we explore the part,
proportional to $\Lambda$, we have $N_i\equiv0$. Now the system ($\r{f}$) gives
\be
\partial_a\Lambda^i(0)-\partial_i\Lambda_a(0)=0
\ee
Thus, if $\Lambda$ does not satisfy this restriction,
one can not restore the gauge invariance without changing the
action ($\r{111}$).

\end{document}